\documentclass[12pt]{iopart}

\usepackage{epsfig}
\usepackage{iopams}

\newcommand{\ket}[1]{\left | #1 \right\rangle}
\newcommand{\bra}[1]{\left \langle #1 \right |}

\renewcommand{\epsilon}{\varepsilon}

\begin{document}

\title[Reference frames for Bell inequality violation in the presence
of superselection rules]{Reference frames for Bell inequality
  violation in the presence of superselection rules}

\author{T Paterek$^{1}$, 
P Kurzy\'nski$^{1,2}$,
D K L Oi$^{1,3}$,
and D Kaszlikowski$^{1,4}$\footnote{Corresponding author.}}

\address{
  $^1$ Centre for Quantum Technologies, National University of Singapore, 3 Science Drive 2, 117543 Singapore, Singapore \\
  $^2$ Faculty of Physics, Adam Mickiewicz University, Umultowska 85, 61-614 Pozna\'{n}, Poland \\
  $^3$ SUPA Department of Physics, University of Strathclyde, Glasgow G4 0NG, United Kingdom \\
  $^4$ Department of Physics, National University of Singapore, 2 Science Drive 3, 117542 Singapore, Singapore}

\ead{phykd@nus.edu.sg}

\begin{abstract}
  Superselection rules (SSRs) constrain the allowed states and operations in
  quantum theory.  They limit preparations and measurements hence
  impact upon our ability to observe non-locality, in particular the
  violation of Bell inequalities.  We show that a reference frame
  compatible with a particle number SSR does not allow
  observers to violate a Bell inequality if and only if it is prepared
  using only local operations and classical communication. In
  particular, jointly prepared separable reference frames are
  sufficient for obtaining violations of a Bell inequality. We study the
  size and non-local properties of such reference frames using
  superselection-induced variance. These results suggest the need for
  experimental Bell tests in the presence of superselection.
\end{abstract}

\maketitle

\section{Introduction}

Symmetries impose powerful constraints in physics, leading Wick
\textit{et al.} to suggest that the associated conserved quantities
lead to additional restrictions on quantum theory, the so-called
superselection rules (SSRs)~\cite{WWW}.  They conjectured that
superselection prevents the existence of coherent superpositions of
charge, for example.  However, Aharonov and Susskind~\cite{AS} showed
that the ability to observe superpositions depends on having a shared
reference frame relative to which the system can be prepared and
measured.  More generally, elements of quantum theory require
reformulation in the presence of SSRs.  Quantum entanglement and
various forms of non-locality are particular examples of phenomena
that are affected by the presence of SSRs, and a vast body of literature already
exists on these topics~\cite{VC}-\cite{BABICHEV}.

Here, we focus on Bell inequality violation in the presence of SSRs.
We concentrate on the bi-partite case where \emph{both} the entangled
principal system and any ancilla/reference frame are subject to
particle-number superselection.  In particular, we examine the role
of the measurement apparatus or reference frame used by the two
observers (our prototypical Alice and Bob). We find that reference
frames prepared using only local operations which satisfy SSRs and
classical communication (SSR-LOCC) \emph{cannot} reveal the
non-locality of an entangled system.  However, jointly prepared but
separable reference frames can be used to violate a Bell inequality
with an entangled principal system.  By imposing separability of the
reference frame, we deduce that violation is due to the measured
entangled state; the reference allows the observers to carry out
measurements that lead to a violation of the Bell inequality. In such
cases, the reference is said to activate Bell violation.

Previous work has explored the issue of Bell inequality violation in
the presence of SSRs, given suitable reference
frames~\cite{AV}-\cite{HA2009}. In contrast to some previous works, here the reference frame is both
explicitly separable and compliant with the SSRs. We show in
general that all references prepared jointly, and only such
references, can activate Bell violation. We find minimal reference
frames and relate the degree of Bell violation to the `non-locality'
in the reference as measured by superselection-induced
variance~\cite{SVC,SVC2}.  This holds in particular for measurements
of single particle states and is a clear proof that such states can
exhibit non-locality
\cite{TAN}-\cite{AMN2007_2}.
We also discuss related single-photon
experiments~\cite{HESSMO,BABICHEV} and conclude that there is still
the need for new experiments.

\section{Scenario}

\begin{figure}[!b]
\begin{center}
\qquad \qquad
\includegraphics{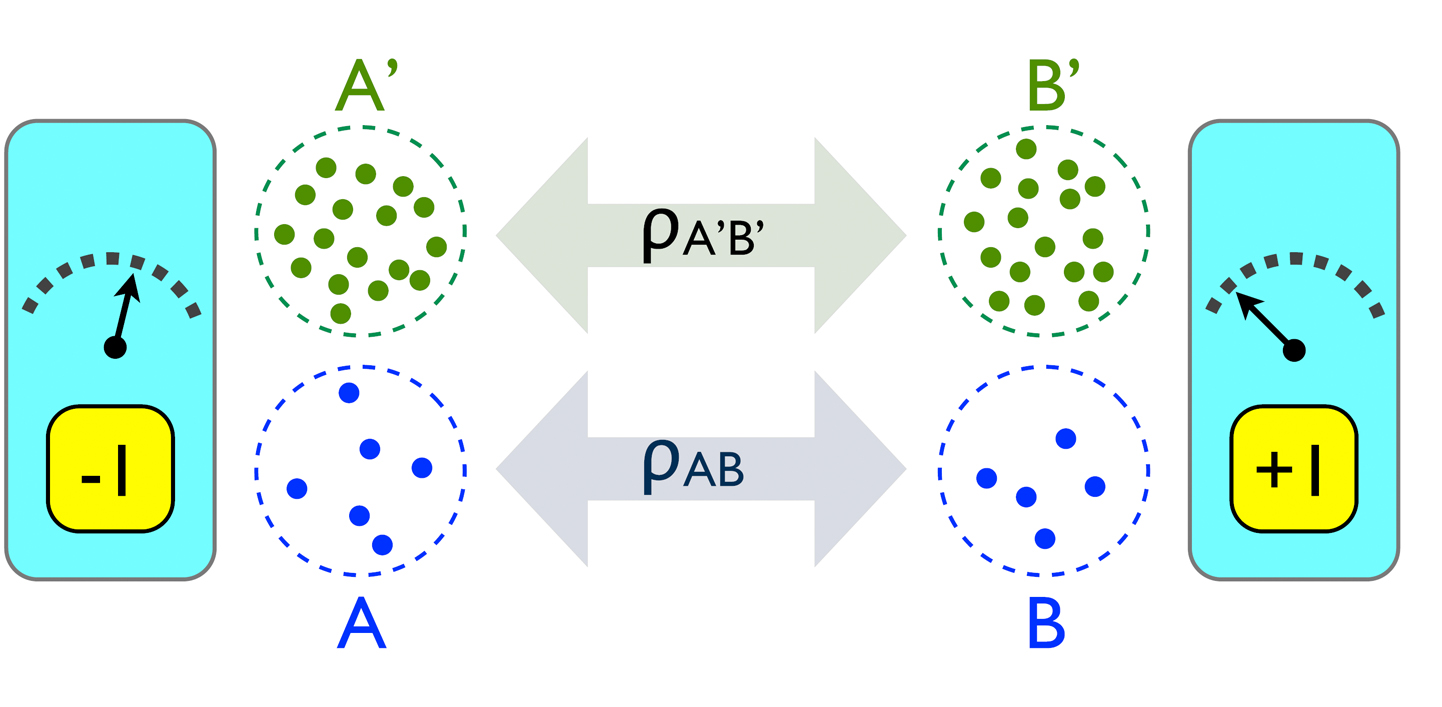}
\end{center}
\caption{Bell experiment in the presence of particle-number
SSR.  Two observers share in advance a reference
  system in a separable state $\rho_{A'B'}$.  The experiment begins
  with an emission of a principal system in a state $\rho_{AB}$.
  According to the SSR both $\rho_{A'B'}$ and
  $\rho_{AB}$ are incoherent mixtures of states with well-defined
  total number of particles.  Alice (on the left) has now access to
  subsystems $A$ and $A'$ and similarly Bob (on the right) has access
  to $B$ and $B'$.  Both local subsystems are next measured by
  superselection-constrained observables, described by projections
  onto states with well-defined total number of particles in the
  subsystems.  In a given experimental run, one of many observables is
  measured at each site, the choice of which is depicted by a tuneable
  knob (arrow) on the measuring device.  Finally, one of many
  measurement results is obtained as depicted by a number on the
  yellow screen (e.g. $-1$ for Alice, $+1$ for Bob).  In short, for
  this scenario a violation of a Bell inequality can be observed if and
  only if the reference state $\rho_{A'B'}$ is prepared jointly, and
  that the more particles in the reference state the larger the
  violation.}
 \label{FIG_BELL}
\end{figure}

We begin with the description of a Bell experiment in the presence of
particle-number superselection. Consider the situation as shown in
Fig.~\ref{FIG_BELL}.  A source distributes to Alice and Bob an
entangled pure state of $N$-particles,
\begin{equation}
\ket{\psi}_{AB} = \sum_{n=0}^N c_n \ket{n}_A \ket{N-n}_B,
\label{PRINCIPAL}
\end{equation}
where $|c_n|^2$ is the probability that Alice finds she has $n$ particles.
Under particle-number superselection, all states and measurements
commute with the particle-number operator $\hat{N}$.  Ordinarily, all
Alice and Bob can do is simply count the number of particles each
receives.  In order to do more than this, they share in advance a joint reference
frame in the state $\rho_{A'B'}$, assumed to be separable and
also obeying particle-number superselection.
Therefore, before state (\ref{PRINCIPAL}) is distributed to Alice and Bob they share no entanglement.
By making joint SSR respecting measurements on each of their respective halves of the entangled system and reference frame ($\{A,A'\}$ for Alice, $\{B,B'\}$
for Bob), they hope to be able to demonstrate a Bell inequality
violation.

\section{Reference frames prepared locally}

We first show that reference frames prepared using only local
operations satisfying SSRs and classical communication (SSR-LOCC)
cannot activate violation of any Bell inequality. The proof is
straightforward.

All such reference states commute with \emph{local} particle-number
operators and therefore are of the form
\begin{equation}
\rho_{A'B'}^{\mathrm{SSR-LOCC}} = \sum_{k,l} p_{kl} \ket{k}_{A'} \! \bra{k} \otimes \ket{l}_{B'} \! \bra{l},
\label{SSR-LOCC-REF}
\end{equation}
where $k$ ($l$) counts particles in the reference frame of Alice
(Bob).  These states contain only classical correlations between fixed
local number of particles as measured by quantum discord and similar quantities
\cite{DISCORD1}-\cite{DISSONANCE}.

Consider for the moment that the reference frame is in the pure state
$\ket{k}_{A'} \ket{l}_{B'}$.  We can express the initial joint state
of the system and reference frame as $\sum_{n} c_n \ket{n,k}_{AA'}
\ket{N-n,l}_{BB'}$ grouping subsystems accessible to Alice and Bob,
respectively.  Note that every term of this superposition contains a
different number of local particles, i.e. $n+k$ for Alice and $N-n+l$
for Bob.  As local SSR observant measurements project onto states with
a well-defined number of local particles, only one term in the
superposition can contribute to the probability of a corresponding
result.  This, however, is exactly the same as making the
measurements on a state in a mixture of $\ket{n,k}_{AA'}
\ket{N-n,l}_{BB'}$ with probability $|c_n|^2$, and this separable
state clearly admits a local hidden variable model.  If one replaces
$|c_n|^2$ in this model with $|c_n|^2 p_{kl}$ all measurement results
obtained with a general mixed reference state (\ref{SSR-LOCC-REF}) are
reproduced. Therefore, no Bell inequality can be violated.

\section{Reference frames prepared jointly}

We now show that all reference frames which cannot be prepared via
SSR-LOCC are useful for Bell violation. We begin with the
characterization of such references. We can express all reference
states in the general form
\begin{equation}
\rho_{A'B'} = \sum_{N'} p_{N'} \rho_{N'},
\end{equation}
where $p_{N'}$ is the probability of $N'$ particles in the reference
frame and $\rho_{N'}$ is any state with a fixed total number of
particles $N'$, i.e. $\rho_{A'B'}$ is an arbitrary mixture of pure
states of the form $\ket{\phi}_{A'B'} = \sum_{i=0}^{N'} r_i
\ket{i}_{A'}\ket{N'-i}_{B'}$.  Since we assume that $\rho_{A'B'}$
cannot be prepared via SSR-LOCC, it necessarily contains off-diagonal
elements in the particle number basis. All such states have a
non-vanishing expectation of
\begin{equation}
\mathcal{V} = \textrm{Re}[\textrm{Tr}(R_+ \otimes R_- \rho_{A'B'})],
\label{COHERENCE}
\end{equation}
where $R_+ = \sum_{a=0}^{N-\Delta} \ket{a+ \Delta} \bra{a}$ and $R_- =
\sum_{b=\Delta}^N \ket{b-\Delta} \bra{b}$ for some $\Delta \ge 1$.  To
see this, note that $\mathcal{V}$ is proportional to the average value
of the sum $\sum_{i=0}^{N'-\Delta} r_{i+\Delta}^* r_{i}$ over the pure
states $\ket{\phi}_{A'B'}$ in the decomposition of $\rho_{A'B'}$.
Therefore, $\mathcal{V}$ vanishes if the sums vanish for all the pure
states.  For states $\ket{\phi}_{A'B'}$ that have coherences in the
particle number basis, this only occurs if the signs of the products
$r_{i+\Delta}^* r_{i}$ alternate for some values of $i$ leading to
cancellation in the sum.  In such a case however, we can always choose
a larger $\Delta$ to skip the terms which lead to the cancellation.
Thus, a state $\rho_{A'B'}$ can be prepared via SSR-LOCC if and only if
$\mathcal{V} = 0$ for all values of $\Delta \ge 1$.

As we now show, all reference states with non-vanishing coherence
parameter $\mathcal{V}$ enable observers to choose measurements that
lead to a violation of the Clauser--Horne-Shimony-Holt (CHSH)
inequality \cite{CHSH}.  Consider an entangled principal system in the
state
\begin{equation}
\ket{\psi}_{AB} = \frac{1}{\sqrt{2}}(\ket{2}_A \ket{2 + \Delta}_B + \ket{2 + \Delta}_A \ket{2}_B).
\end{equation}
We show in Appendix A that there always exist dichotomic local
measurements compatible with the SSR whose outcomes on the joint state
of the principal system and reference frame are correlated as
\begin{equation}
E(\alpha_k, \beta_l) = - \cos(2 \alpha_k) \cos(2 \beta_l) + \mathcal{V} \sin(2 \alpha_k) \sin(2 \beta_l),
\label{CORR}
\end{equation}
where the angle $\alpha_k$ ($\beta_l$) parameterizes the $k$th ($l$th)
setting of Alice (Bob).  We insert this expression into the CHSH parameter
\begin{equation}
S \equiv E(\alpha_1,\beta_1) + E(\alpha_1,\beta_2) + E(\alpha_2,\beta_1) - E(\alpha_2,\beta_2),
\end{equation}
and find values of $\alpha_k$ and $\beta_l$ for which $S$ is higher
than the local realism bound $S\le 2$. Namely, we choose $\alpha_1 = 0$,
$\alpha_2 = \pi/4$, and parameterize the settings of Bob by a single
angle $\beta \equiv \beta_1 = - \beta_2$, leading to
\begin{equation}
S = -2 \cos(2 \beta) + 2 \mathcal{V} \sin(2 \beta).
\end{equation}
To find its maximum, note that $S$ has a form of a scalar product
between the vector $\vec w = (-2,2 \mathcal{V})$ and an arbitrary
normalized vector $\vec v = (\cos(2 \beta), \sin(2 \beta))$.
Therefore, there always exists an angle $\beta$ such that $\vec v$ is
parallel to $\vec w$ and the maximum of the scalar product is given by the
length of $\vec w$:
\begin{equation}
S = 2 \sqrt{1 + \mathcal{V}^2}.
\end{equation}
To summarize, all reference frames that cannot be prepared via
SSR-LOCC have a non-vanishing coherence parameter $\mathcal{V}$ and
consequently allow observers to carry out measurements on entangled
states that lead to a violation of the CHSH inequality:
\begin{equation}
S > 2 \quad \textrm{for all} \quad \mathcal{V} \ne 0.
\end{equation}
An identical conclusion holds for single-particle entangled principal system.
The calculations are the same as long as the reference frame does not contain any vacuum.

\section{Minimal separable reference frames}

We have derived conditions for violating the CHSH inequality in the
presence of particle-number SSR and we further study the properties of
reference frames activating the violation.  We show here
that the minimal \emph{separable} reference frame allowing violation
contains two particles in total.

Relaxing for a moment the separability requirement,
Eq. (\ref{COHERENCE}) shows that the minimal reference has only one
particle in total.  Namely, any state $\ket{\phi}_{A'B'} = r_0
\ket{0}_{A'} \ket{1}_{B'} + r_1 \ket{1}_{A'} \ket{0}_{B'}$ with $r_1
r_0^* \ne 0$ has a non-vanishing parameter $\mathcal{V}$. However, the
application of the PPT criterion for entanglement \cite{PPT1,PPT2}
reveals that $\ket{\phi}_{A'B'}$ is entangled for all $r_1 r_0^* \ne
0$.  When entangled states are used as references, it is unclear whether
the violation of a Bell inequality is due to the entanglement of the
principal system or the reference frame.

For this reason we consider a reference frame with at most two
particles:
\begin{equation}
\rho_{A'B'} = p_{00} \ket{00}_{A'B'} \! \bra{00} + p_{11} \ket{11}_{A'B'} \! \bra{11} + p_\phi \ket{\phi}_{A'B'} \! \bra{\phi},
\label{MINIMAL}
\end{equation}
where as before $\ket{\phi}_{A'B'} = r_0 \ket{0}_{A'} \ket{1}_{B'} +
r_1 \ket{1}_{A'} \ket{0}_{B'}$.  Since the definition of
Eq. (\ref{COHERENCE}) involves only the real part of the off-diagonal
elements, we choose the coefficients $r_0$ and $r_1$ to be real,
i.e. $\mathcal{V} = p_\phi r_0 r_1$, and we have used $\Delta = 1$ in
Eq.~(\ref{COHERENCE}).  The application of the PPT criterion reveals
that the state $\rho_{A'B'}$ is separable if and only if
\begin{equation}
p_{00} p_{11} \ge p_{\phi}^2 r_0^2 r_1^2 = \mathcal{V}^2.
\label{PPT_V}
\end{equation}
Therefore, for all separable reference frames activating the
violation, $\mathcal{V} \ne 0$, there must be some mixture of the
two-particle state ($p_{11} > 0$).  Note that the same argument
applies to $p_{00}$ and one concludes that separable reference states of the form (\ref{MINIMAL})
enabling the violation must contain some vacuum.  This is a
consequence of the fact that an arbitrary mixture of any pure two-qubit
entangled state with `colored noise' $\ket{11} \bra{11}$ is always
entangled~\cite{BADZIAG,SZUMY}.

\section{Local and global twirling}

A useful mathematical tool which illustrates and generalizes the results
presented so far is the twirling operation. Twirling $\mathcal{T}$
eliminates the coherences that are not compatible with SSR:
\begin{equation}
\mathcal{T}(\rho) \equiv \sum_{n} \Pi_n \rho \Pi_n,
\end{equation}
where $\Pi_n$ is a projector on a subspace with a fixed number of
particles $n$.  This operation describes the
lack of a reference frame enabling access to the phase information
of the probability amplitudes.

The usefulness of twirling is best illustrated by considering states
that can be prepared via SSR-LOCC. A simple proof demonstrates that
they cannot activate CHSH violation. A SSR-LOCC reference frame
commutes with local particle number operators and therefore it is invariant
under the action of local twirlings, $\rho_{A'B'}^{\mathrm{SSR-LOCC}}
= (\mathcal{T}_{A'} \otimes
\mathcal{T}_{B'})(\rho_{A'B'}^{\mathrm{SSR-LOCC}})$.  This implies for
the coherence parameter
\begin{eqnarray}
  \mathcal{V} &\sim& \Tr \left\{ R_+\otimes R_-\rho_{A'B'}^{\mathrm{SSR-LOCC}} \right\} \nonumber \\
  & = &\Tr \left\{ (R_+\otimes R_- )(\mathcal{T}_{A'} \otimes \mathcal{T}_{B'})\rho_{A'B'}^{\mathrm{SSR-LOCC}} \right\} \nonumber \\
  & = &\Tr\left\{\mathcal{T}_{A'}(R_+) \otimes \mathcal{T}_{B'}(R_-)\rho_{A'B'}^{\mathrm{SSR-LOCC}} \right\} =0,
  \label{VIS}
\end{eqnarray}
where the last equality follows from the fact that
$\mathcal{T}_{A'/B'}(R_\pm) = 0$ because $R_{\pm}$ contain only
off-diagonal elements in the particle-number basis.

Note that the operator $R_+ \otimes R_-$ conserves total particle number and
therefore it is invariant under global twirling, $\mathcal{T}(R_+
\otimes R_-) = R_+ \otimes R_-$. All states satisfying SSR are of the
form $\rho_{A'B'}^{\mathrm{SSR}} = \mathcal{T} (\rho_{A'B'})$, where
now $\rho_{A'B'}$ need not have a fixed number of particles.
Therefore,
\begin{eqnarray}
\Tr\{ R_+\otimes R_- \rho_{A'B'}^{\mathrm{SSR}} \} & = & \Tr \{ R_+\otimes R_- \mathcal{T}(\rho_{A'B'}) \} = \Tr \{ \mathcal{T}(R_+\otimes R_-) \rho_{A'B'} \} \nonumber \\
& = & \Tr \{ R_+\otimes R_- \rho_{A'B'} \},
\end{eqnarray}
where we have used the cyclic property of trace. This means that in order
to calculate $\mathcal{V}$ for a SSR respecting reference frame
$\rho_{A'B'}^{\mathrm{SSR}}$, we can use in Eq. (\ref{COHERENCE}) any
state $\rho_{A'B'}$ whose twirling gives $\rho_{A'B'}^{\mathrm{SSR}}$.

\section{Separable reference for maximal violation}

Twirling allows further study of separable reference frames. Note that
the minimal reference frame of two particles we have derived in
Eq.~(\ref{MINIMAL}) activates the violation but does not lead to
maximal violation. Indeed, the highest value of $\mathcal{V}$ for
entangled reference states (\ref{MINIMAL}) is $\frac{1}{2}$ and for
separable reference states Eq.~(\ref{PPT_V}) implies $\mathcal{V} \le
\frac{1}{4}$, whereas the maximal violation of the CHSH inequality
allowed by quantum theory $S=2\sqrt{2}$ \cite{TSIRELSON} requires
$\mathcal{V} = 1$.  Note also that the Tsirelson bound of
$S=2\sqrt{2}$ implies that $|\mathcal{V}| \le 1$.

Here we show that there are separable reference frames allowing
maximal violation of the CHSH inequality with an entangle state.
First note that all separable states satisfying SSR are of the form
$\rho_{\mathrm{sep}}^{\mathrm{SSR}} = \mathcal{T}
(\rho_{\mathrm{sep}})$, with $\rho_{\mathrm{sep}} = \sum_j p_j
\ket{a_j'} \bra{a_j'} \otimes \ket{b_j'} \bra{b_j'}$ where $\ket{a_j'}
\ket{b_j'}$ need not have a fixed number of particles.  This follows
from the fact that global twirling is an LOCC operation (but not
SSR-LOCC) and as such cannot produce entanglement.
We now derive the separable reference frames that maximize the coherence parameter $\mathcal{V}$.
Since Eq. (\ref{COHERENCE}) is linear in $\rho_{A'B'}$, $\mathcal{V}$ is
maximal for a pure product state $\ket{a'} \ket{b'}$.  Moreover, due
to the fact that only the real part enters (\ref{COHERENCE}), it is
sufficient to consider states with real coefficients $\ket{a'} =
\sum_{n=0}^{N} \mathfrak{a}_n\ket{n}$ and $\ket{b'} = \sum_{m=0}^{M}
\mathfrak{b}_m\ket{m}$ with $\mathfrak{a}_n, \mathfrak{b}_m \in
\mathbb{R}$.  For such pure states $\mathcal{V} = f_{N} g_{M}$ where
\begin{eqnarray}
f_{N} & \equiv & \bra{a'} R_+ \ket{a'} =  \sum_{n=0}^{N-1} \mathfrak{a}_n \mathfrak{a}_{n+1}, \\
g_{M} & \equiv & \bra{b'} R_- \ket{b'} =  \sum_{m=0}^{M-1} \mathfrak{b}_{m} \mathfrak{b}_{m+1},
\end{eqnarray}
with $N$ and $M$ denoting the dimensionality of the reference of Alice
and Bob respectively, i.e. the maximal number of particles in their
reference frames, and we put $\Delta = 1$. To find the maximum of
$\mathcal{V}$, it is now sufficient to optimize $f_N$, because $g_M$
has the same form and optimization over Bob's state is independent of
that over Alice's.  Note that for states with real coefficients
$\bra{a'} R_+ \ket{a'} = \bra{a'} R_- \ket{a'}$ and therefore $f_N =
\frac{1}{2}\bra{a'} (R_+ + R_-) \ket{a'}$. The only non-vanishing
elements of matrix $R_+ + R_-$ are a strip of identities above and
below its diagonal, and therefore it is a Hermitian matrix.  The maximal
value of $f_N$ is attained for $\ket{a'}$ being the eigenvector of
$R_+ + R_-$ with the highest eigenvalue.  The amplitudes of the
optimal state read
\begin{equation}
\mathfrak{a}_n = \sqrt{\frac{2}{N+2}} \sin \left(\frac{\pi (n+1)}{N+2} \right), \quad \textrm{with} \quad n=0,1,\dots,N,
\end{equation}
and its maximal eigenvalue gives
\begin{equation}
\max f_N = \cos \left( \frac{\pi}{N+2} \right).
\label{MAX_FN}
\end{equation}
For small references containing at most one particle on both sides
$N=M=1$ we find $\mathcal{V} \le \frac{1}{4}$ in agreement with the
results of section $5$ on minimal reference frames.  If the references of Alice
and Bob are both unbounded, Eq.~(\ref{MAX_FN}) implies that
$\mathcal{V}\to 1$ and the violation of the CHSH inequality approaches
its maximum.  Practically, for $N = M \approx 30$ particles in each
reference frame, $\mathcal{V} \approx 0.99$.

The references for violation of Bell inequality were also
studied in~\cite{BRF} in the context of directional reference frames,
finding that in the limit of unbounded reference frame, maximal violation
can be achieved.  We stress that in our case the corresponding limit
is twofold: to maximally violate CHSH inequality the reference has to
be prepared jointly, and it should be unbounded.

\section{Non-locality of reference states}

Let us now discuss the relation between ${\cal V}$ and the
non-locality of reference frames as captured by SIV~\cite{SVC}.  We show that violation of the CHSH
inequality is a witness of non-zero SIV in the reference frame and
that the amount of SIV in small references gives an upper bound on the
CHSH violation.

The SIV of a pure state $\ket{\phi}$ is defined as the variance of local
number of particles
\begin{equation}
\label{SIV}
\frac{1}{4} V(\phi) \equiv \langle\phi|N_A^2\otimes I |\phi\rangle-\langle\phi|N_A \otimes I |\phi\rangle^2.
\end{equation}
The factor of $4$ is introduced for normalization: one unit of SIV is
defined for the state
$\frac{1}{\sqrt{2}}(|n,n+1\rangle+|n+1,n\rangle)$.  Since SIV is
symmetric with respect to permutation of the parties, one can equally
consider the variance of the local particle number on Bob's side
($N_B$). Pure states that have non-zero SIV cannot be prepared via
SSR-LOCC and this is the property of reference frames we are
interested in.  However, such pure states are always entangled, whereas
we insist on separability of the reference frame. Therefore we must
consider mixed states. Just like entanglement, for mixed
states another measure of SIV is required. We shall use the
variance of formation defined as~\cite{SVC}
\begin{equation}
V_F^{SSR}(\rho)=\min_{\{p_i,\phi_i\}}\sum_i p_i V(\phi_i),
\end{equation}
where the ensembles of pure states $\{\phi_i\}$ obey SSRs. As a
measure of the off-diagonal terms in the density matrix, we can
consider $\mathcal{V}$, or equivalently Bell inequality violation, as
a witness of non-zero SIV. Moreover, it was shown in Ref. \cite{SVC2}
that for states (\ref{MINIMAL}) the variance of formation reads
$V_F^{SSR}(\rho) \geq 4 p_{\phi}^2 r_0^2 r_1^2 = 4 \mathcal{V}^2$
with real $r_0$ and $r_1$.  Accordingly, one can directly relate SIV
to $\cal V$ as
\begin{equation}
|{\cal V}|\leq \frac{\sqrt{V_F^{SSR}(\rho)}}{2}.
\end{equation} 
Therefore, the corresponding states with vanishing SIV have also vanishing $\mathcal{V}$.

\section{Experiments}

Our last topic is the experimental verification of Bell inequality
violations under SSRs and the need for new experiments. We relate this
by commenting on current experiments related to the Bell inequality
and single-photon non-locality~\cite{TAN,HESSMO,BABICHEV}.  Although not
intended to violate a Bell inequality under an SSR, these experiments
may be seen as such for (an induced) photon-number SSR
\cite{REF_REVIEW}. In the proposal of Ref.~\cite{TAN}, a single photon
is directed onto a balanced beam-splitter producing (it is hoped) a
non-local state of one photon. In each output port of this first
beam-splitter there is another balanced beam-splitter with a
(reference) coherent state directed at its second input port (see
Fig.~1 of Ref.~\cite{TAN}).  One considers correlations between the
number of photons registered in detectors placed in the output ports
of the second set of beam-splitters.

The experimental realizations~\cite{HESSMO,BABICHEV} differ from the
proposal~\cite{TAN} in that the secondary beam-splitters may be unbalanced.  Note
that in principle the possible measurement results are unbounded, and
therefore the CHSH inequality cannot be applied.  It turns out that
the correlation functions violate the CHSH inequality only for small
mean number of photons in the coherent states, in which case the
events of having many photons in the detectors are rare and the CHSH
inequality becomes applicable.  This, however, opens up an effective
detection loophole which allows for a local hidden variable model.

Let us denote by $r_a$ ($r_b$) reflectivity of the first (second)
beam-splitter supplied with a coherent state.  The corresponding
transmittances are: $t_n = 1 - r_n$ with $n=a,b$.  It is assumed that
both coherent states have the same mean number of photons $\bar n$ and
relative phase $\varphi = \alpha - \beta$.  The correlation function
between the number of photons measured behind the two beam-splitters
reads
\begin{equation}
E_{\varphi} = \frac{(r_a - t_a) (r_b - t_b)(\bar n - 1)+ 4 \sqrt{r_a r_b
t_a t_b}  \sin\varphi }{\bar n+1}.
\end{equation}
Using this expression in the CHSH parameter, one finds that the
proposal of Ref.~\cite{TAN} is optimal in the sense that it is best to
choose balanced beam-splitters $r_a = r_b = \frac{1}{2}$.  Any other
values of $r_{a}$ and $r_b$ lead to smaller values of the CHSH
parameter.  In particular, the assumption of Ref.~\cite{HESSMO} that
after the beam-splitter a photon may have equal likelihood to have
come from a single photon `beam' or a coherent state, i.e. $r \bar n
= t$ leads to no violation for all values of $\bar n$.  For the
balanced beam-splitters the inequality is violated if $\bar n <
\sqrt{2} - 1$, which translates into the critical probability of
vacuum in the coherent state $P_{\mathrm{vac}} \approx \frac{2}{3}$.
Using such coherent states it is quite rare to measure two photons in
a setup and one may utilize this effective detection loophole to
explain the observed results with, e.g., the model of
Gisin~\cite{GISINS}.

We therefore hope this research will stimulate further experiments
testing Bell violation in the presence of SSRs.  Ideally one would use
systems with natural SSRs such as massive particles or
charges. However, studies of partial superselection can also be
performed through controlled decoherence, as decoherence between
different subspaces can be seen as a type of SSR.

\section{Conclusions}

We have studied the effects of restrictions imposed by SSRs on Bell inequality violation.  We found that the violation
primarily depends on how a reference frame is prepared and only
secondarily on its size.  Even unbounded reference frames do not lead
to Bell violation if they are prepared via SSR-LOCC and therefore are strictly classically correlated according to quantum discord and similar measures.
This condition
was shown to be necessary and sufficient for the violation;
that is reference frames enable the violation of a Bell inequality if and
only if they cannot be prepared via SSR-LOCC.  The violation can be
achieved with separable reference frames explicitly consistent with
particle-number superselection, the minimal such reference containing
up to two particles.  We linked the violation to the amount of
non-locality in the reference frame as captured by SIV. It would be interesting to study how other
subfields of quantum theory, e.g. quantum tomography, are modified in
the presence of superselection.

\section{Acknowledgements} 

We acknowledge discussions with {\v C}. Brukner, B. Hessmo and
J. Vaccaro.  We are grateful to an anonymous referee for pointing out
a problem in the previous version of the manuscript which eventually led us
to stronger results.  This research is supported by the National
Research Foundation and Ministry of Education in Singapore.  DKLO
acknowledges the support of the Quantum Information Scotland network
(QUISCO).

\appendix

\section{Derivation of correlations formula (\ref{CORR})}

Assume for the moment that the reference system is in a pure state
\begin{equation}
\ket{\phi}_{A'B'} = \sum_{i=0}^{N'} r_i \ket{i}_{A'} \ket{N'-i}_{B'}.
\end{equation}
Since it cannot be prepared via SSR-LOCC, we have $r_{i + \Delta}^*
r_i \ne 0$ for some $i$ and $\Delta \ge 1$.

Consider principal system in a state $\ket{\psi} = \frac{1}{\sqrt{2}}
(\ket{2}_A \ket{2 + \Delta}_B + \ket{2 + \Delta}_A \ket{2}_B)$ such
that the initial state of the principal system and reference together
reads:
\begin{equation}
\ket{\psi \phi} = \sum_{i} \frac{r_i}{\sqrt{2}} (\ket{2,i}_{AA'} \ket{2 + \Delta, N'-i}_{BB'} + \ket{2+\Delta,i}_{AA'} \ket{2,N'-i}_{BB'}),
\end{equation}
where we grouped in kets subsystems accessible to Alice and Bob
respectively.

We present local dichotomic measurements compatible with particle-number SSR which lead to the correlation function (\ref{CORR}) of the
main text. Alice measures a local observable parameterized by angle
$\alpha$:
\begin{equation}
\mathcal{A} = \sum_{a = - \Delta}^{N'} \ket{\alpha(a)} \bra{\alpha(a)} - \sum_{a = - \Delta}^{N'} \ket{\bar \alpha(a)} \bra{\bar \alpha(a)},
\end{equation}
where the eigenstates are defined as follows:
\begin{eqnarray}
\ket{\alpha(a)} & = & \cos \alpha \ket{1,a+\Delta+1}_{AA'} + \sin\alpha \ket{2,a+\Delta}_{AA'}, \\
\ket{\bar \alpha(a)} & = & \sin\alpha \ket{1,a+\Delta+1}_{AA'} - \cos\alpha \ket{2,a+\Delta}_{AA'},
\end{eqnarray}
for $a = - \Delta, \dots, -1$;
\begin{eqnarray}
\ket{\alpha(a)} & = & \cos \alpha \ket{2 + \Delta, a}_{AA'} + \sin\alpha \ket{2,a+\Delta}_{AA'}, \\
\ket{\bar \alpha(a)} & = & \sin\alpha \ket{2+\Delta,a}_{AA'} - \cos\alpha \ket{2,a+\Delta}_{AA'},
\end{eqnarray}
for $a = 0, \dots, N'-\Delta$;
\begin{eqnarray}
\ket{\alpha(a)} & = & \cos \alpha \ket{2 + \Delta, a}_{AA'} + \sin\alpha \ket{3+\Delta,a-1}_{AA'}, \\
\ket{\bar \alpha(a)} & = & \sin\alpha \ket{2+\Delta,a}_{AA'} - \cos\alpha \ket{3+\Delta,a-1}_{AA'},
\end{eqnarray}
for $a = N'-\Delta+1, \dots, N'$.  These observables are compatible
with the SSR because all the eigenstates contain a fixed total number of
particles $2 + a + \Delta$.  The reason behind the three different
cases is that the reference subsystem cannot contain more than $N'$
particles and less than zero.  They are also chosen to form an
orthonormal set of vectors.  To obtain the observables of Bob one just
needs to replace $A \to B$, $A' \to B'$, $\alpha \to \beta$ and $a \to
b$.

We reverse the equations for the eigenvectors and write the initial
state of the principal system and the reference as
\begin{eqnarray}
\ket{\psi \phi} & = & \sum_{i} \frac{r_i}{\sqrt{2}} \Big\{ (\sin\alpha \ket{\alpha_{i-\Delta}} - \cos\alpha \ket{\bar \alpha_{i-\Delta}} ) (\cos\beta \ket{\beta_{N'-i}} + \sin\beta \ket{\bar \beta_{N'-i}}) \nonumber \\
& + & (\cos\alpha \ket{\alpha_i} + \sin\alpha \ket{\bar \alpha_i}) (\sin\beta \ket{\beta_{N'-i-\Delta}} - \cos\beta \ket{\bar \beta_{N'-i-\Delta}}) \Big\}.
\end{eqnarray}
The probabilities of the results corresponding to different eigenvectors are:
\begin{eqnarray}
P_{ab} & \equiv & |\langle \alpha(a) \beta(b)| \psi \phi \rangle|^2 = \frac{1}{2}|r_{a+\Delta} \sin\alpha \cos\beta +r_a \cos \alpha \sin \beta|^2 \delta_{b,N'-a-\Delta} \nonumber \\
P_{a \bar b} & \equiv & |\langle \alpha(a) \bar \beta(b)| \psi \phi \rangle|^2 = \frac{1}{2}|r_{a+\Delta} \sin\alpha \sin\beta - r_a \cos \alpha \cos\beta|^2 \delta_{b,N'-a-\Delta} \nonumber \\
P_{\bar a b} & \equiv & |\langle \bar \alpha(a) \beta(b)| \psi \phi \rangle|^2 = \frac{1}{2}|- r_{a+\Delta} \cos\alpha \cos\beta +r_a \sin \alpha \sin \beta|^2 \delta_{b,N'-a-\Delta} \nonumber \\
P_{\bar a \bar b} & \equiv & |\langle \bar \alpha(a) \bar \beta(b)| \psi \phi \rangle|^2 = \frac{1}{2}| - r_{a+\Delta} \cos\alpha \sin\beta - r_a \sin \alpha \cos \beta|^2 \delta_{b,N'-a-\Delta} \nonumber
\end{eqnarray}
Note that for every $a$ there is only one $b$ for which the
corresponding probability does not vanish, and it is easy to verify
that indeed $\sum_{a,b = - \Delta}^{N'} (P_{ab} + P_{a \bar b} +
P_{\bar ab} + P_{\bar a \bar b}) = 1$.  Finally, the correlation
function is the average of the product of dichotomic local results
\begin{equation}
E_{\phi}(\alpha,\beta) = \sum_{a,b = - \Delta}^N(P_{ab} + P_{\bar a \bar b} - P_{\bar ab} - P_{a \bar b}).
\end{equation}
Plugging in the formulae for the probabilities, we obtain
\begin{equation}
E_\phi(\alpha,\beta) = - \cos(2 \alpha) \cos(2 \beta) + \mathcal{V} \sin(2 \alpha) \sin(2 \beta),
\label{APP_CORR}
\end{equation}
where $\mathcal{V} = \sum_{a = 0}^{N'- \Delta} \mathrm{Re}(r_{a+
  \Delta}^* r_a)$.  Alternatively $\mathcal{V}$ can be expressed using
operators $R_+ = \sum_{a=0}^{N'-\Delta} \ket{a + \Delta} \bra{a}$ and
$R_- = \sum_{b=\Delta}^{N'} \ket{b - \Delta} \bra{b}$ with the help of
which $\mathcal{V} = \mathrm{Re}(\bra{\phi} R_+ \otimes R_-
\ket{\phi})$.  This calculation holds for \emph{arbitrary} pure state
$\ket{\phi}_{A'B'}$ of the reference.  Therefore, for the reference in
an arbitrary mixed state $\rho_{A'B'} = \sum_{\phi}p_\phi
\ket{\phi}_{A'B'} \! \bra{\phi}$ the correlations formula reads
\begin{equation}
E(\alpha,\beta) = \sum_\phi p_\phi E_\phi(\alpha,\beta),
\label{APP_E_MIX}
\end{equation}
and therefore it is of the same form as Eq. (\ref{APP_CORR}), but with
the modified coherence parameter
\begin{equation}
\mathcal{V} = \mathrm{Re}[\Tr(R_+ \otimes R_- \rho_{A'B'}) ].
\end{equation}

\end{document}